\documentclass[notitlepage,floatfix,onecolumn,aps,showpacs,superscriptaddress,longbibliography,10]{revtex4-1}

\usepackage{graphicx}
\usepackage{multirow,booktabs}
\usepackage{amsmath,amssymb,latexsym,MnSymbol}
\usepackage[colorlinks=true, urlcolor=blue,linkcolor=blue,citecolor=blue]{hyperref}

\makeatletter
\let\cat@comma@active\@empty
\makeatother

\usepackage{lineno}
\usepackage{verbatim}

\begin{document}

\title{Centrality anomalies  in complex networks as a result of model over-simplification}

\author{Luiz G. A. Alves}
\email{lgaalves@northwestern.edu}
\affiliation{Department of Chemical and Biological Engineering, Northwestern University, Evanston, IL 60208, USA}
\affiliation{Institute of Mathematics and Computer Science, University of S\~ao Paulo, S\~ao Carlos, SP 13566-590, Brazil}

\author{Alberto Aleta}
\affiliation{Department of Theoretical Physics, University of Zaragoza, Zaragoza 50009, Spain}
\affiliation{Institute for Biocomputation and Physics of Complex Systems (BIFI), University of Zaragoza, Zaragoza 50009, Spain}

\author{Francisco A. Rodrigues}
\affiliation{Institute of Mathematics and Computer Science, University of S\~ao Paulo, S\~ao Carlos, SP 13566-590, Brazil}
\affiliation{Mathematics Institute, University of Warwick, Gibbet Hill Road, Coventry CV4 7AL, UK}
\affiliation{Centre for Complexity Science, University of Warwick, Coventry CV4 7AL, UK}

\author{Yamir Moreno}\email{yamir.moreno@gmail.com}
\affiliation{Department of Theoretical Physics, University of Zaragoza, Zaragoza 50009, Spain}
\affiliation{Institute for Biocomputation and Physics of Complex Systems (BIFI), University of Zaragoza, Zaragoza 50009, Spain}
\affiliation{ISI Foundation, Turin 10126, Italy}

\author{Lu\'is A. Nunes Amaral}\email{amaral@northwestern.edu}
\affiliation{Department of Chemical and Biological Engineering, Northwestern University, Evanston, IL 60208, USA}
\affiliation{Department of Physics and Astronomy, Northwestern University, Evanston, IL 60208, USA}
\affiliation{Northwestern Institute on Complex Systems (NICO), Northwestern University, Evanston, IL 60208, USA}

\begin{abstract}
Tremendous advances have been made in our understanding of the properties and evolution of complex networks. These advances were initially driven by information-poor empirical networks and theoretical analysis of unweighted and undirected graphs. Recently, information-rich empirical data complex networks supported the development of more sophisticated models that include edge directionality and weight properties, and multiple layers. Many studies still focus on unweighted undirected description of networks, prompting an essential question: how to identify when a model is simpler than it must be? Here, we argue that the presence of centrality anomalies in complex networks is a result of model over-simplification. Specifically, we investigate the well-known anomaly in betweenness centrality for transportation networks, according to which highly connected nodes are not necessarily the most central. Using a broad class of network models with weights and spatial constraints and four large datasets of transportation networks, we show that the unweighted projection of the structure of these networks can exhibit a significant fraction of anomalous nodes compared to a random null model. However, the weighted projection of these networks, compared with an appropriated null model, significantly reduces the fraction of anomalies observed, suggesting that centrality anomalies are a symptom of model over-simplification. Because lack of information-rich data is a common challenge when dealing with complex networks and can cause anomalies that misestimate the role of nodes in the system, we argue that sufficiently sophisticated models be used when anomalies are detected. 
\end{abstract}

\maketitle
\section*{Introduction}
The study of complex networks produced fruitful results in many areas of knowledge, from systems biology~\cite{guimera2005functional,park2013structural} and social systems~\cite{girvan2002community,wang2013interdependent} to epidemiology~\cite{cohen2010complex,helbing2015saving,moreno2002epidemic} and statistical physics~\cite{pastor2003statistical,newman2018networks}. The initial focus of complex networks and graph theory was on undirected, unweighted topologies~\cite{barrat2008dynamical,newman2018networks}. Using unweighted network projections, many properties were proved to be effective in describing complex systems~\cite{strogatz2001exploring,newman2003structure,Amaral2004,boccaletti2006complex}. More recently, weighted, directed, multiplexed networks have been the focus of much research attention. In many cases, these more sophisticated representations of the system are most appropriate to describe real-world networks~\cite{Barrat2004,Buldyrev2010,kivela2014multilayer,Boccaletti2014}. Despite it, researchers still fall back on representing a system's network of interactions as if it was undirected and unweighted, many times because of the lack of information-rich datasets. 

This is the case of gene regulatory networks, where usually direction, strengths, and signs of the links are overlooked because of the lack of complete data~\cite{sanz2012topological}. Another case where empirical studies have overlooked the details of the system is the case of multipartite networks~\cite{benson2018simplicial}. This class of systems comprises networks with multiple groups that can only interact through nodes of different types. However, because of the lack of information-rich datasets, these systems are usually studied after projection onto networks of one single type of node. Thus, the question is how to determine when such a model is good enough to represent the system, especially in the absence of data for testing simulation predictions.

Here, we focus on the case of weighted networks projected onto unweighted networks. We propose that the presence of anomalies in the structure of the undirected and unweighted projection of the network can be a result of a situation where a model is simpler than it must be.  Our starting observation is the report of betweenness centrality anomalies in transportation networks~\cite{guimera2005worldwide}.  This simple measure can capture the importance of a node to connect different parts of the network~\cite{newman2018networks} by the means of how often it stands between other nodes. Guimer\`a et al. reported that nodes with a large degree in air transportation networks do not necessarily have the highest betweenness centrality, whereas some low degree nodes can have large betweenness centralities.  The emergence of these anomalies has been attributed to the multi-community structure of the network and spatial constraints such as geopolitical boundaries~\cite{guimera2005worldwide,barrat2005effects,barthelemy2011spatial}. Nevertheless, the general mechanisms governing the emergence of such anomalies remain unknown.

In order to tackle these questions, we investigate a broad class of network models with weights and spatial constraints and the structure of four transportation networks. Our analysis reveals that, like for the class of model networks, unweighted transportation networks exhibit centrality anomalies for a significant fraction of the nodes compared with an appropriate null model with the same degree distribution. However, these anomalies disappear when we consider weighted representations of the network. Our findings support the hypothesis that such centrality anomalies are a symptom of a model that is simpler than it must be.

Because model over-simplification might lead to anomalies that would misestimate the role of nodes in the system, our findings have direct implications for the modeling of dynamical processes on complex networks where betweenness centrality is used to measure the influence of nodes, such as in the modeling of human dynamics~\cite{barbosa2018human}, the spread of diseases~\cite{meloni2009traffic,meloni2011modeling}, crime spreading~\cite{caminha2017human}, and spatial networks~\cite{barrat2005effects,barthelemy2011spatial}. Moreover, they also hint at the significant challenges when modeling biological~\cite{sanz2012topological}, economic, or social phenomena because data incompleteness is so pervasive.

\section*{Results}
\subsection*{Centrality anomalies}
\begin{figure}
\begin{center}
\includegraphics[width=\columnwidth]{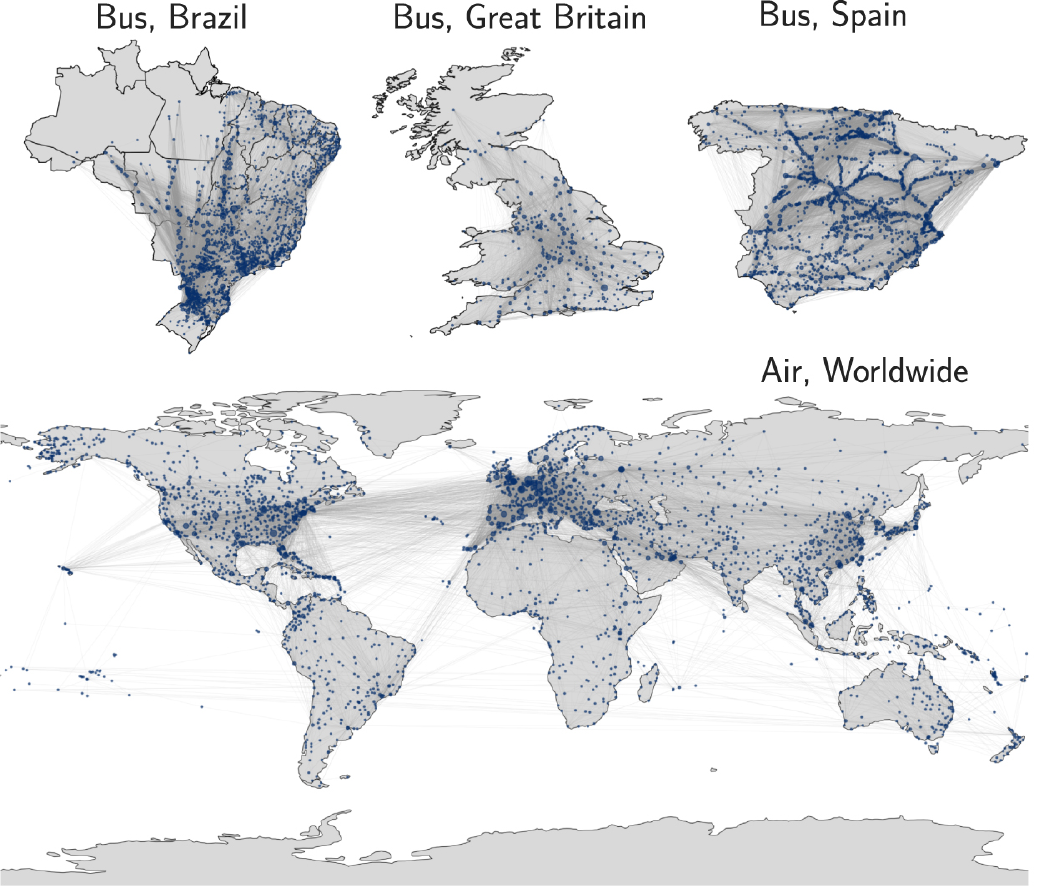} 
\end{center}
\caption{Illustration of the transportation network datasets. We collected four large datasets of transportation networks that include information about the number of buses or airplanes on each route. The data consist of the inter-city bus transportation networks of three countries (Brazil, Great Britain, and Spain), and worldwide air transportation network. In the plots, the node area is proportional to the degree of the node.}
\label{fig:maps}
\end{figure}

We collected extensive data for four large scale transportation networks: Brazil, Great-Britain, and Spain bus transportation networks, and the worldwide air transportation network.  We define an inter-city bus transportation network by assigning a node to each of the $N$ municipalities (with at least one bus station) and assigning an undirected edge between two nodes if the two stops $i$ and $j$ are connected by at least one bus route. Throughout the period observed for each data set, the same route can be offered by more than one company and multiple times by a single company (see methods for details). This fact enables us to define the weight of the edge, $w_{ij}$, as the total number of buses offered by all companies over the observation period (Fig.~\ref{fig:maps}).

In the worldwide air transportation network, each node represents a city. As a consequence, if there are multiple airports serving the same city, we assign the relevant airports to a single node. For example, JFK, La Guardia, and Newark airports are all assigned to the New York City node. We assigned undirected edges between two nodes $i$ and $j$ if the two cities were connected by at least one air route. Because not all air routes have daily or greater frequency, and in order not to drop less-traveled cities, we collected information on flights occurring during the week of May 17, 2018, to May 22, 2018. As for the bus transportation networks, the same route can be offered by more than one company and multiple times a day by the same company. Thus, we defined the weight of an edge, $w_{ij}$, as the total number of flights offered by different companies flying the route during the observation period (Fig.~\ref{fig:maps}).

\begin{figure}
\begin{center}
\includegraphics[width=\columnwidth]{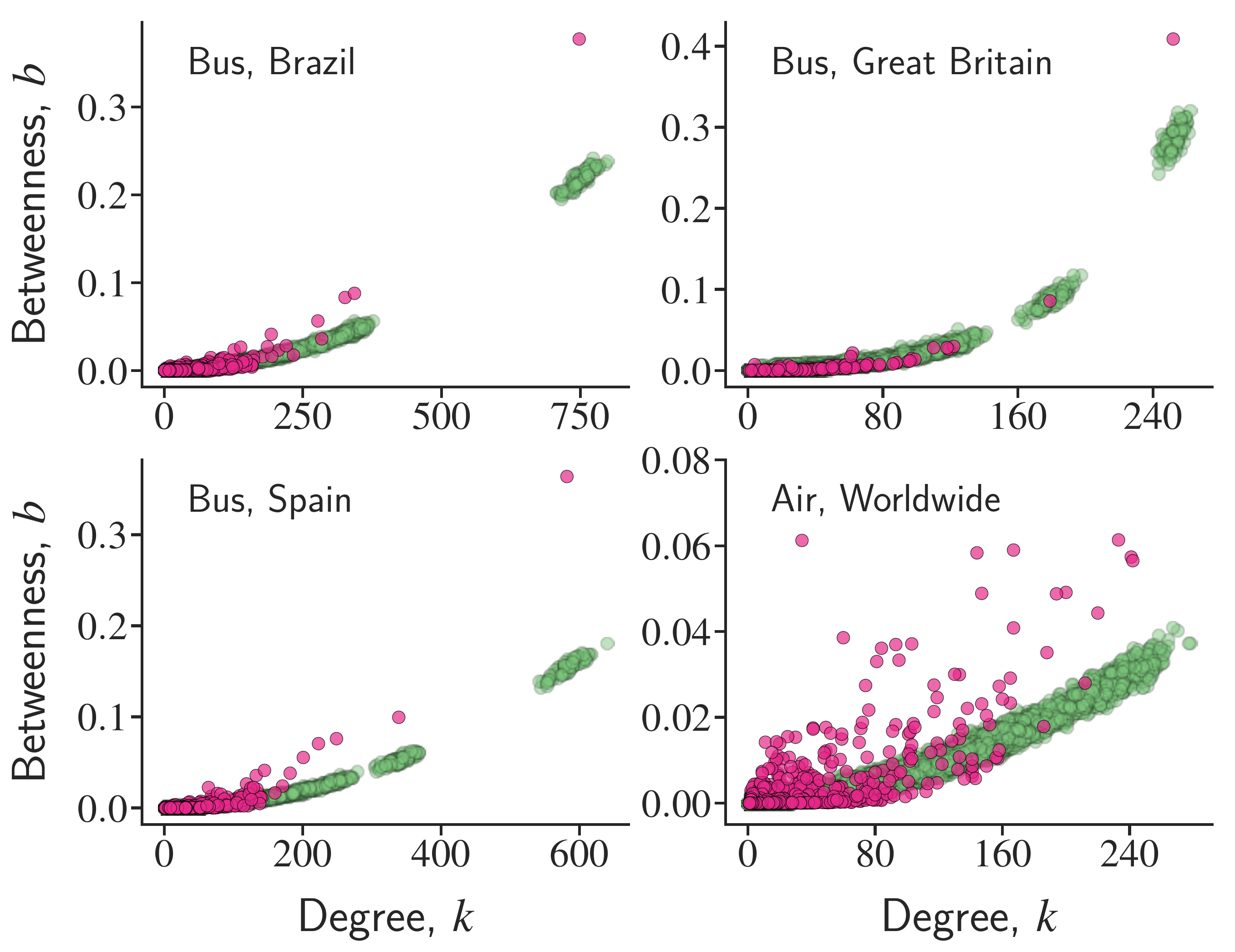} 
\end{center}
\caption{Investigation of centrality anomalies for the unweighted transportation networks. The panels show the comparison of real data (pink circles) with an ensemble of 10,000 networks (green circles) generated using the UBCM method~\cite{squartini2015unbiased}.  As expected, for the randomized networks the betweenness centrality scales approximately with the degree squared. In contrast, for the empirical networks, the relationship between degree and betweenness is much less straightforward as there are some nodes for which the betweenness dramatically deviates from the scaling relationship.}
\label{fig:bvsk_ubcm}
\end{figure}

Several studies have reported that spatial networks, such as the ones we study here,  can exhibit centrality anomalies~\cite{guimera2004modeling,guimera2005worldwide,barrat2005effects,mukherjee2012statistical} --- that is, the betweenness centrality of a node is not necessarily proportional to its degree squared. First, we investigate to what extent these centrality anomalies are due to the over-simplification of the networks.  Specifically, we first calculate the betweenness centrality $b$ and degree $k$ of the nodes for an unweighted projection of the network. The betweenness centrality of node $i$ counts the fraction of shortest paths connecting all pairs of nodes that pass through node $i$ but do not include node $i$~\cite{freeman1977set}. Fig.~\ref{fig:bvsk_ubcm} shows the betweenness centrality versus degree for the networks studied here. 

In order to make sense of the observed values of the betweenness and their relationship with the degree, we compare the measurements for the four transportation networks to the expected values for ensembles of randomized networks with the same degree distributions. In order to provide consistency with later analyses, we do not use the typical Markov chain Monte Carlo edge switching approach, in which the structural constraints are satisfied exactly (i.e., {\it microcanonical ensemble}), and instead implement the undirected binary configuration model (UBCM)~\cite{squartini2015unbiased}, where the constraints are met on average over the ensemble (i.e., {\it canonical ensemble})~\cite{bianconi2007entropy,squartini2011analytical,gabrielli2018grand}. In the UBCM, edges are placed at random following a distribution that preserves, on average, the original degree distribution observed in the data (see methods). 

As has been reported earlier~\cite{guimera2004modeling,guimera2005worldwide}, the betweennesses obtained for the randomized networks do not recapitulate those observed for the empirical networks. That is, whereas there is an approximate scaling of the betweenness with the degree squared for the randomized networks, for the empirical networks one finds many nodes with large deviations from that scaling relationship.  

\subsection*{Model networks}

It has been proposed that the existence of these centrality anomalies is due to the presence of spatial constraints and the special role, due to economic or political considerations, that some cities might have~\cite{guimera2004modeling,barthelemy2011spatial,barrat2005effects}. However, the precise factors driving the emergence of such anomalies remain unknown. 

To investigate the generality of our findings, we next study a class of spatial weighted networks generated using the {\it Strength Driven Preferential Attachment with Spatial Selection} (SDPASS) model, which has been reported to produce centrality anomalies~\cite{barrat2005effects}. In this model, $N_0$ initial nodes are randomly located on a two-dimensional disc of radius $L$ according to a uniform distribution and they are connected by links with weights $w_0$. At each time-step, a new node $i$ is placed randomly on the disc, following a uniform distribution. The new node is connected to $m$ previously existent nodes that are preferentially near and have the largest strength, according to, 
\begin{equation}
p_{ij}= \frac{s_j\, e^{-d_{ij}/r_c}}{\sum_l s_l\, e^{-d_{il}/r_c}}
\end{equation}
\noindent where $r_c$ is a desired spatial scale, $s_i$ is the strength of the node (i.e. $s_i=\sum_j w_{ij}$), and $d_{li}$ is the Euclidean distance between nodes $l$ and $i$. The new edge $(i,j)$ has a fixed weight $w_0$ and the creation of this edge perturbs the existing links attached to node $j$. To add this local perturbation to the model, the weights between $j$ and its neighbors $l \in \mathcal{V}(j)$ are modified following the rule:
\begin{equation}
w_{jl}\rightarrow w_{jl} + \delta \frac{w_{jl}}{s_j},
\end{equation}
where $\delta$ characterizes the susceptibility of the network to new links and $s_j=\sum_k w_{jk}$ is the strength of node $j$. If $\delta < w_0$, the new link has a small influence on the network. If $\delta \approx w_0$, the newly created traffic on the new edge is transferred to existing connections. If $\delta > w_0$, the traffic in the new edge generates a multiplicative effect on the traffic of the neighbors. This process is repeated until the network reaches the desired size. It is worth to note that this process generates a symmetric adjacency matrix, i.e. $w_{ij}=w_{ji}$, a necessary condition for the null models we use.

We explore the SDPASS model for networks with $N_0=5$ initial nodes, $m=4$, and size $N=100$. We simulate all relevant limiting cases to explore how $\delta$ and the ratio $\eta=r_c/L$ affects the scaling of the betweenness centrality. For convenience, we fixed $L=1$ to explicitly explore the dependence of the model on $r_c$. For each set of parameters, we generated a network using the SDPASS model, and, subsequently, we used the appropriated null models to generate an ensemble of networks to calculate the fraction of anomalous nodes in these networks.

To make the identification of centrality anomalies rigorous, we compare the observed values of the pair $(k_i,b_i)$ of node $i$ to the distribution of expected values for the randomized ensemble. We find that the distribution of expected values is reasonably approximated by a multivariate Gaussian, $\mathcal{N}(\mathbf x | \boldsymbol\mu_i,\boldsymbol\Sigma_i)$, where $\boldsymbol\mu_i$ represents the average values of $k_i$ and $b_i$ for the random ensemble and $\boldsymbol\Sigma_i$ represents the covariance matrix. We fit a multivariate Gaussian to the random ensemble data for each node and use it to compute the line enclosing $95\%$ of the probability mass (see methods for details). 

Considering $\eta \gg 1$ the effects of distance are negligible~\cite{barrat2005effects} and we recover the non-spatial weighted network model of Barrat {\it et al.}~\cite{barrat2004weighted}, which showed no anomalies in our simulations compared with an ensemble of networks generated by the UBCM model. As $\delta \rightarrow 0$, the weight effects are no longer significant and we recover the preferential attachment model~\cite{simkin2011re}. The preferential attachment model does not show any anomalies in the betweenness centrality, and an ensemble of random networks generated by the UBCM model is able to predict the betweenness centrality of the nodes. For instance, using $\delta=0.01$ and $\eta=10$ and comparing this network with an ensemble of networks generated by the UBCM model we found that only $1\%$ of the nodes have centrality anomalies.

Another possible scenario is $\delta \ll 1$ and  $\eta \ll 1$. In this case, the effect of the link's weights is negligible and we essentially have a spatial unweighted network topology. In this case, the centrality anomalies are also not present, and our random network model (UBCM) is able to predict the betweenness centrality of the nodes. Using $\delta=0.01$ and $\eta=0.01$ to generate our network and comparing it with an ensemble of networks that preserves the degree distribution (UBCM), we found that only $1\%$ of the nodes are anomalous.


Finally, we investigate the interplay between weights dynamics, i.e, $\delta \geq 1$, and spatial constraints, $\eta \ll 1$. In these limits, the model generates spatial weighted networks that have centrality anomalies similar to the ones observed for transportation networks. For instance, using $\delta=10$ and $\eta=0.01$, we found a significant fraction of nodes ($\approx 69\%$) that show anomalies in the unweighted projection of the network when compared to the ensemble of networks produced by the UBCM model.

\begin{figure}
\begin{center}
\includegraphics[width=\columnwidth]{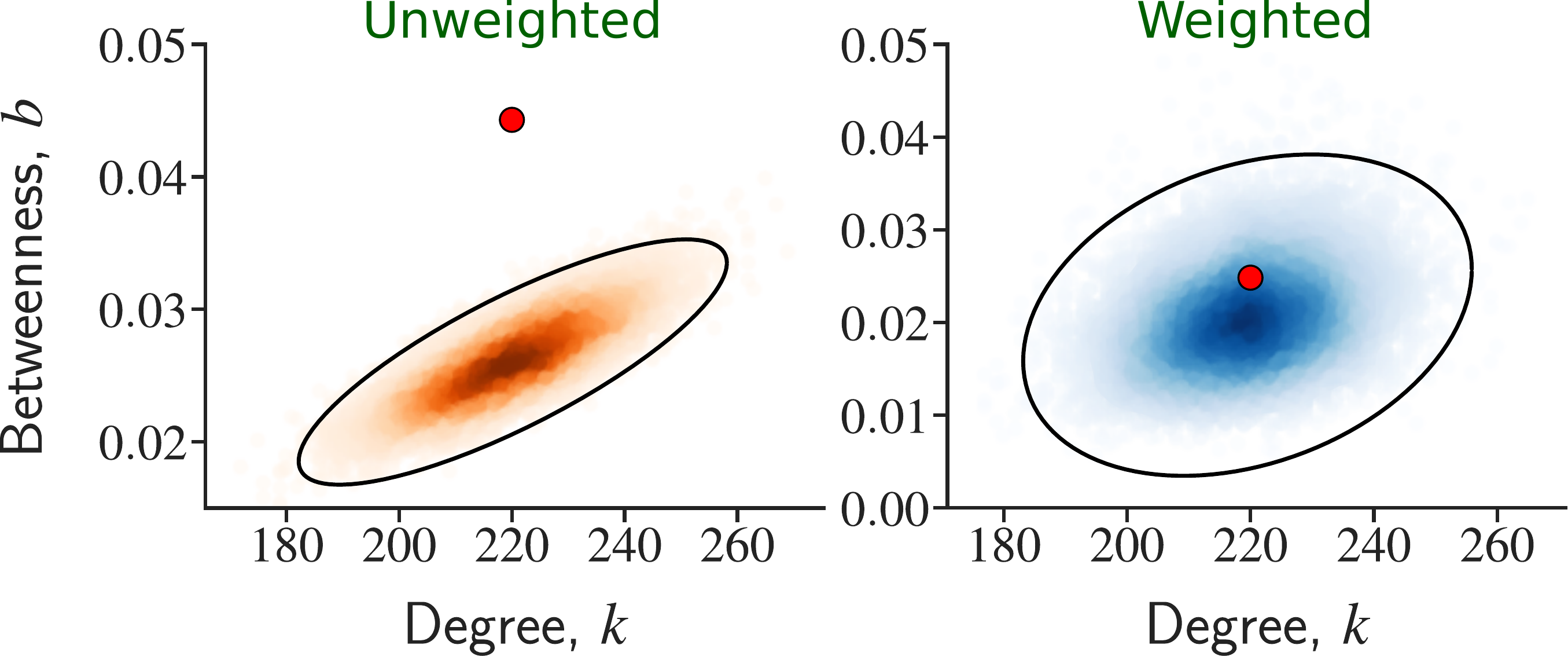} 
\end{center}
\caption{Illustration of centrality anomalies identification in complex networks. The red dot indicates the observed centrality and degree empirical network and the orange (blue) dots are the corresponding betweenness centrality vs degree for the $10,000$ networks from the ensemble sampled using the UBCM (UECM) method. The solid black line encloses $95\%$ of the probability mas for a multivariate Gaussian fit to the data. In the unweighted network, the observed values of betweenness centrality and degree lie outside the $95\%$ bounds of the multivariate Gaussian adjusted to the data predicted by the synthetically generated networks. In contrast, in the weighted network, an anomaly is no longer observed.}
\label{fig:anomalousdefinition}
\end{figure}

Next, we compare the measurements for the model network to the expected values for an ensemble of randomized networks with the same degree and strength distributions. To this end, we use the undirected enhanced configuration model (UECM)~\cite{mastrandrea2014enhanced,squartini2015unbiased}, which, consistently with the UBCM, preserves the constraints on average over the ensemble (i.e., {\it canonical ensemble})~\cite{bianconi2007entropy,squartini2011analytical,gabrielli2018grand}. In the UECM, edges and their weights are placed at random following distributions that, on average, preserve both the degree and the strength of the nodes; see methods. Note that the weights $w_{ij}$ in our empirical networks represent the number of buses or airplanes available for the route connecting $i$ and $j$. While higher values of $w$ do reflect stronger ties, a physically appropriate calculation of the path length requires that one quantifies the length of an edge as the inverse of its weight~\cite{brandes2008variants}.  Consistently with the transportation networks, we next consider the inverse of the weights to compute betweenness centrality of our model network.  In Fig.~\ref{fig:anomalousdefinition} we show for illustration purposes the betweenness centrality data for both the unweighted and weighted randomizations.  It is visually apparent that there is a centrality anomaly for one case but not the other.

Using the weighted projection of our model network and comparing it with an ensemble of networks generated by the UECM model, the fraction of centrality anomalies decrease to $18\%$ of the nodes, a much smaller fraction than the $69\%$ detected for the unweighted projection. Note that, because our null model does not include spatial information, our results suggest that a more sophisticated model would be a better choice for representing this network. The results of our model networks are summarized in Table~\ref{table:1}.

\begin{table}
\begin{tabular}{rrrrr}
{\bf $\delta$ } & {\bf $\eta$ } & {\bf UBCM}  & {\bf  UECM} & {\bf Topology } \\
\hline
0.01 & 10 & 1\%  & 1\% & Non-spatial unweighted \\
10 & 10 & 1\%  & 1\% & Non-spatial weighted\\
0.01 & 0.01  & 1\% & 1\%  & Spatial unweighted  \\
10 & 0.01  & 69\% & 18\%  &  Spatial weighted  \\

\end{tabular}
\caption{Anomalies in the SDPASS model. The first two columns show the parameters used on a simulation of networks considering $N_0=5$ initial nodes, $m=4$, and size $N=100$. The third column (unweighted network) and fourth column (weighted network) show the percentage of anomalous nodes in these networks when compared with an ensemble of networks generated by the UBCM and UECM models, respectively. The last column indicates the topology characteristics of the networks given the parameters $\delta$ and $\eta$.}\label{table:1}
\end{table}

\subsection*{Weighted transportation networks}
\begin{figure}
\begin{center}
\includegraphics[width=\columnwidth]{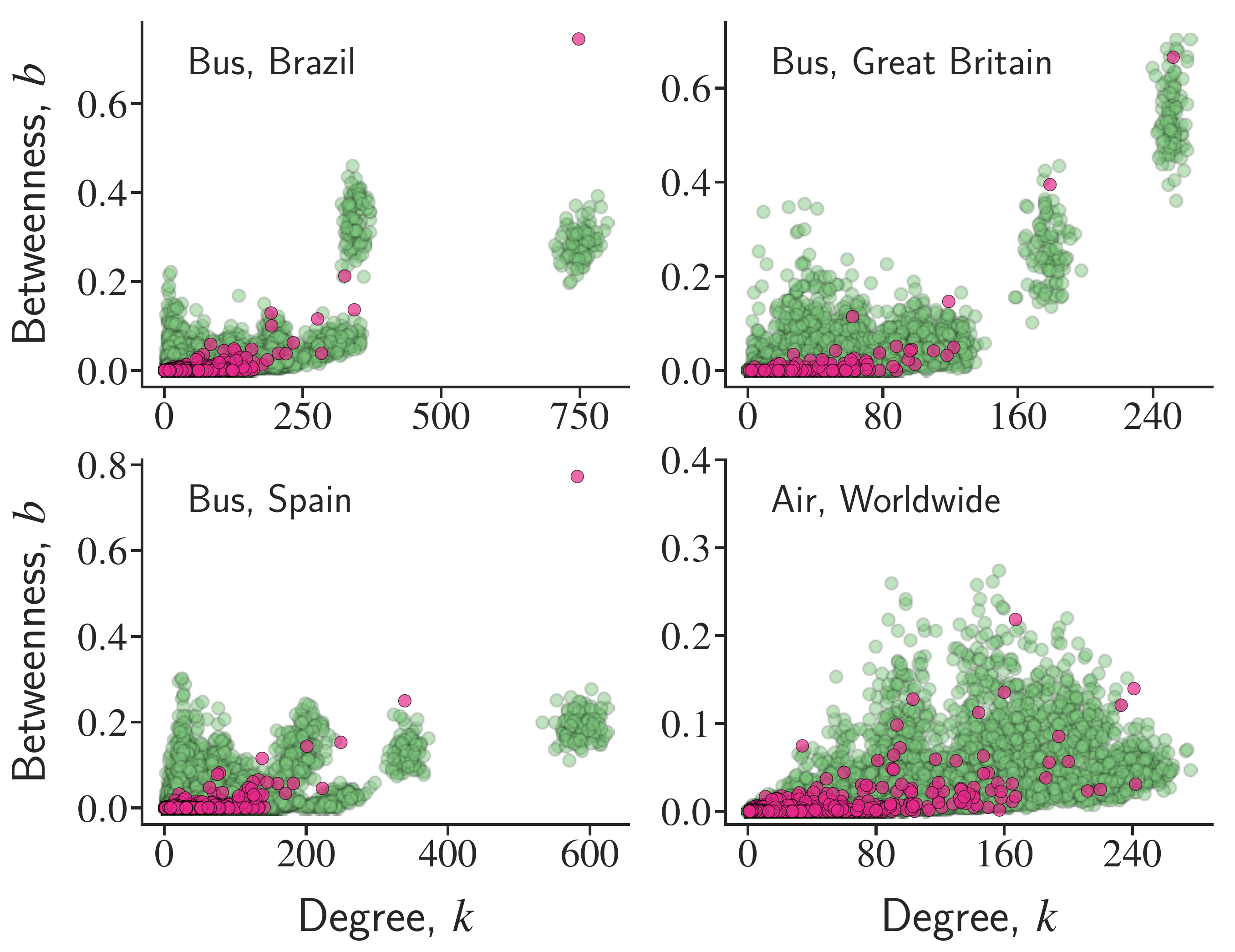} 
\end{center}
\caption{Investigation of centrality anomalies for the weighted transportation networks. The panels shows the comparison of real data (pink circles) with an ensemble of 10,000 networks (green circles) generated using the UECM method~\cite{squartini2015unbiased}. Notice that there no longer exists a simple scaling relationship between betweenness and degree. }
\label{fig:bvsk_uecm}
\end{figure}

\begin{figure}
\begin{center}
\includegraphics[width=\columnwidth]{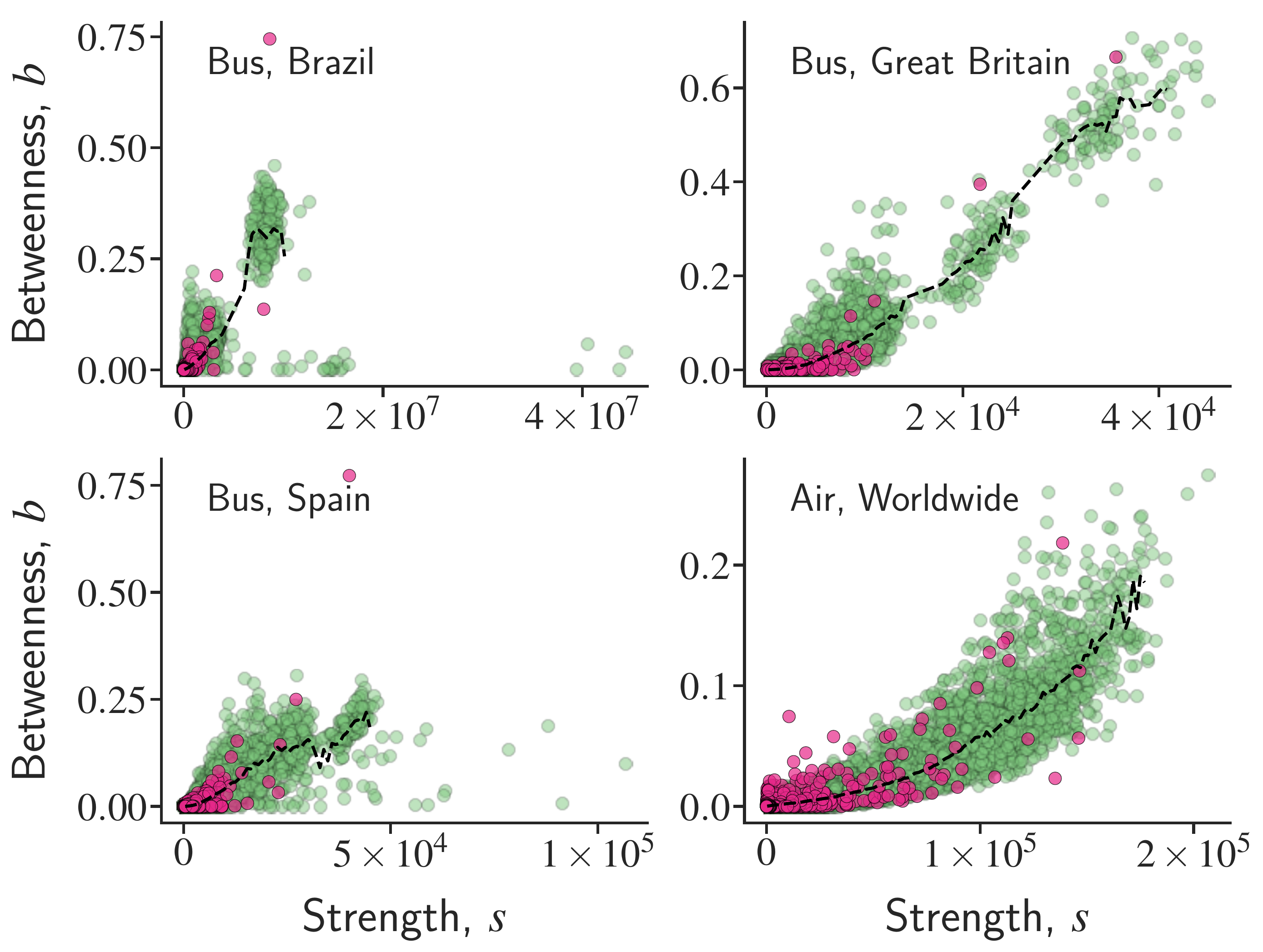} 
\end{center}
\caption{Investigation of betweenness centrality as a function of the strength. The pink circles show observed values, whereas the green circles show results for an ensemble of 10,000 networks generated using the UECM method~\cite{squartini2015unbiased}. The dashed black line is an average over the ensemble data. The average trend suggests that the strength is a more informative measure of the nodes than degree.   }
\label{fig:bvss_uecm}
\end{figure}

To investigate the relevance of the results for networks in the real world systems, we next explore whether centrality anomalies are also present when considering the weighted representation of the transportation networks. As before, we compare the relationship between observed betweennesses and degrees to the relationship obtained for an ensemble of 10,000 randomized networks generated using the UECM (Fig.~\ref{fig:bvsk_uecm}). By doing so, we observe two results. First, even for the randomized networks, there no longer exists a simple scaling relationship between betweenness and degree. Second, we no longer find systematic centrality anomalies in the data.  Remarkably,  only a handful of cities --- Brasilia, Madrid, and Barcelona --- appear to have a centrality anomaly and none of the nodes with low degree appears to have such anomalies. On the other hand, by plotting betweenness vs strength (Fig.~\ref{fig:bvss_uecm}), we uncover a simpler relationship, indicating that the strength would be a more informative measure of the nodes.  

\begin{figure}
\begin{center}
\includegraphics[width=\columnwidth]{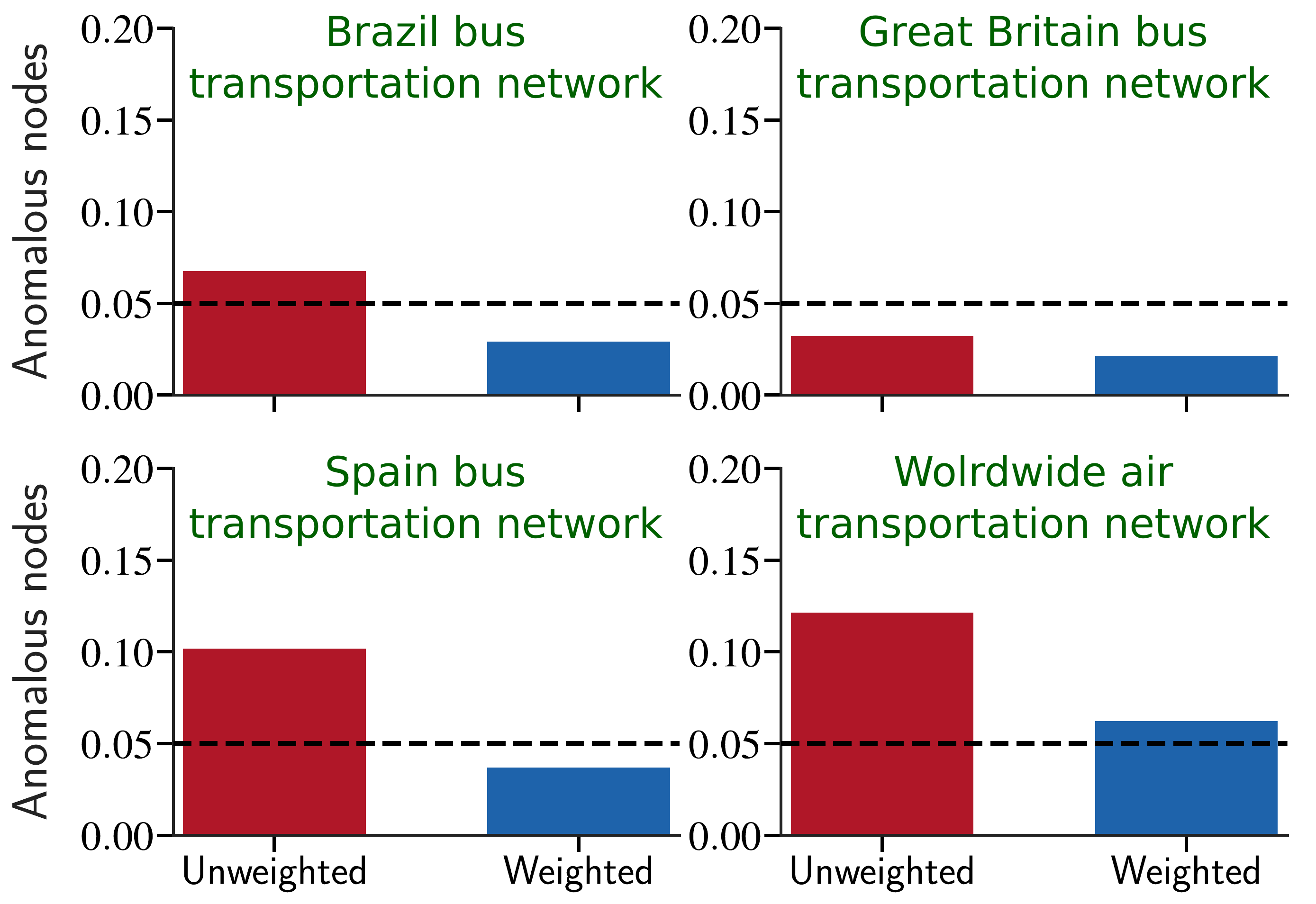} 
\end{center}
\caption{Quantifying the fraction of anomalous nodes. For each network, we compute the fraction of nodes that lie outside the $95\%$ bounds of the model. The anomaly in betweenness centrality is verified for all unweighted transportation networks (except for Great Britain). In contrast, for the weighted version, the fraction of anomalous nodes is of the order of the false discovery rate, i.e., approximately $5\%$.}
\label{fig:anomalousfraction}
\end{figure}

We now calculate the fraction of nodes for which we can reject the null hypothesis of no centrality anomaly (Fig.~\ref{fig:anomalousfraction}). The expectation here is that we will observe a false discovery rate of $5\%$.  For 3 of the 4 unweighted transportation networks, we find an excess of nodes with centrality anomalies, whereas for none of the weighted networks we find such an excess. These results suggest that the existence of centrality anomalies when considering unweighted networks is a result of the neglected (but functionally crucial) role of edge weight on the evolution and performance of these networks.

\section*{Conclusions}

The findings reported here suggest that centrality anomalies present in the unweighted representation of transportation networks are masking the fact that some edges carry much larger weights than the typical edge in the network. Because of the role of spatial, temporal, and capacity constraints in real transportation networks, it is natural to expect that the degree of individual nodes cannot grow unbound, and that edge weight is a way to account for large demand. Indeed, we find that for random networks with the same degree and strength distributions the centrality structure of the network becomes indistinguishable from the observed structure. 

We further extend our results to a broader class of model networks using the {\it strength driven preferential attachment with spatial selection} model. Specifically, we show that when weights and spatial constraints are relevant, the centrality anomalies arise in the unweighted network projection and they cannot be predicted using a simple model that takes into account only the degree sequence as a constraint. On the other hand, when degree and strength sequences are used as a constraint for the null model, the ensemble can reproduce the betweenness centrality observed in the data, suggesting that, in the case of spatial weighted networks, more sophisticated network models are better choices for representing the system.  

Our findings demonstrate that the desire to use the simplest network representations of a system carries important risks. Typically, researchers fall back on models that ignore connection directionality and weight. While this choice may be good enough in many cases, in others it could be masking important characteristics of the system.  Our study shows that the presence of centrality anomalies can be an indicator that important aspects of the system are being lost in its network representation. We believe that complex systems that have nodes and edges embedded in a physical space such as spatial networks (e.g., road networks, power grids, and neural networks), might show centralities anomalies when projected onto unweighted networks. Further investigation of these systems could extend the generality of our findings to other real-world systems.

\section*{Methods}


{\bf Data.} We obtained data from the Brazilian inter-city bus routes for the period between January 2005 to December 2014 at a monthly time-resolution. These data are maintained and distributed by the Brazilian National Land Transportation Agency (ANTT)~\cite{antt}. The data contains more than 19 thousand unique routes connecting 1786 cities. We gathered the geographical location of all relevant cities from the Brazilian Institute of Geography and Statistics (IBGE)~\cite{ibge}. 

We obtained data from the British inter-city bus routes for the period between October 4, 2010, to October 10, 2010, at an hourly resolution. These data are maintained by the National Public Transport Data Repository (NPTDR) and distributed by the Department of Transport and licensed under the Open Government Licence. This dataset was complemented with the National Coach Services Data (NCSD) distributed also by the Department of Transport and licensed under the Open Government Licence \cite{dataUK}. The total number of nodes after the aggregation into municipalities is $279$ comprising almost 4 thousand unique routes.

We obtained data from the Spanish inter-city bus routes for the period between January 1, 2017, to December 31, 2017, at an hourly resolution. These data are maintained and distributed by the Spain Ministry of Development~\cite{dataES}. The data is provided as the set of routes connecting each pair of municipalities in Spain except for the province of Girona. The total number of nodes is $1,435$ with over $20$ thousand unique routes.


The data of the worldwide air transportation network were collected in the period between May 17, 2018, to May 22, 2018, at an hourly resolution. These data are maintained by the website Flight Aware~\cite{FlightAware}. The data contain all flights in 2734 airports around the world, with more than 16 thousand unique routes. The geographical location of the airports was obtained from the Open Flights website~\cite{OpenFlights}.

\begin{figure}[!h]
\begin{center}
\includegraphics[width=\columnwidth]{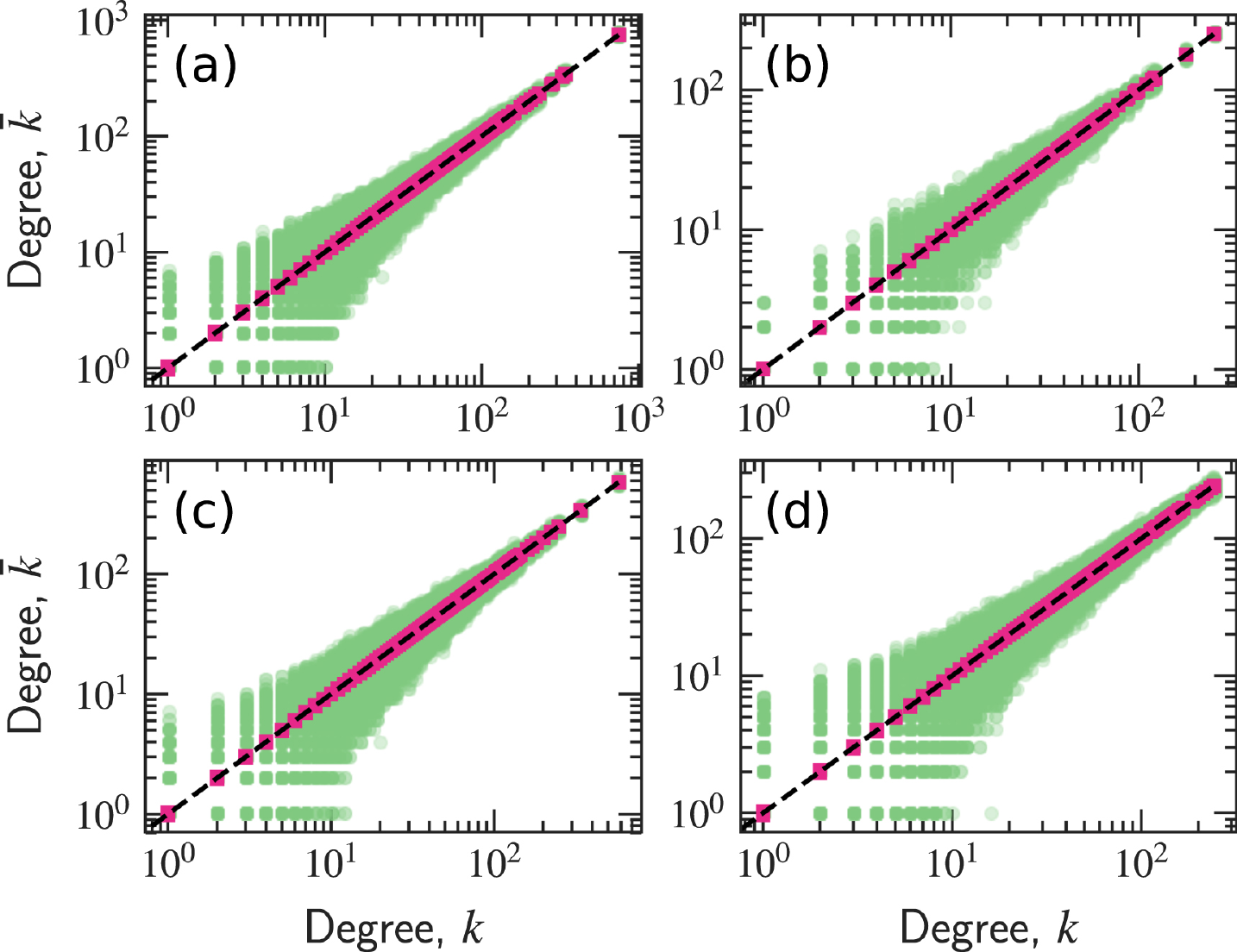} 
\end{center}
\caption{Samples of networks using the undirected binary configuration model (UBCM) given their degree sequence. In each plot, the green dots show the measured degree $\bar{k_i}$ in each sample versus the degree $k_i$ in the observed network. The pink squared-dots represents the average degree $\langle k_i \rangle$ over the $10,000$ networks of the ensemble versus the empirical degree $k_i$. The dashed line is a straight line with slope 1. (a) Brazilian buses transportation network, (b) British buses transportation network, (c) Spanish buses transportation network, and (d) worldwide air transportation network. }
\label{fig:degree_ubcm}
\end{figure}

{\bf Sampling of networks.} To investigate the statistical properties of transportation networks we have generated $10,000$ networks sampled from the ensembles for each dataset and topology (non-weighted or weighted). We followed the approach proposed by Squartini {\it et al.}~\cite{squartini2011analytical,squartini2015unbiased} of unbiased sampling based on maximum-entropy distributions. In this approach, the probability distributions composing the ensemble are obtained by maximizing, in sequence, the Shannon's entropy and the likelihood function subject to the desired constraints. In particular, for the non-weighted networks case we used the ``undirected binary configuration model'' (UBCM), where the constraint is the degree sequence $\{k_i\}_{i=1}^N$. Notice that the constraints in the {\it canonical ensemble} are met on average over the network samples, differently from the {\it microcanonical ensemble}, i.e. Morkov Chain Monte Carlo edge switching approach, where the constraints are satisfied exactly~\cite{bianconi2007entropy,squartini2011analytical,gabrielli2018grand}. With the UBCM model the probability of having a link between nodes $i$ and $j$, $p_{ij}$ is given by
\begin{equation}
p_{ij} \equiv \frac{x_i x_j}{1+x_i x_j}
\end{equation}
\noindent
where the vector ${\bf x}$ of $N$ unknown parameters can be determined by either maximizing the log-likelihood function
\begin{equation}
\lambda({\bf x}) = \sum_i k_i ({\bf A}) \ln x_i - \sum_i \sum_{i<j} \ln (1+x_i x_j)
\end{equation}
\noindent
where ${\bf A}$ refers to the adjacency matrix of the observed graph, or by solving the system of $N$ equations:
\begin{equation}
\langle k_i\rangle= \sum_{j \neq i}\frac{x_i x_j}{1+x_i x_j} = k_i \left({\bf A} \right) \quad \forall i,
\end{equation}

\noindent where $k_i({\bf A})$ is the observed degree of node $i$ and $\langle k_i\rangle$ is the ensemble average. Once the values of the $p_{ij}$ have been determined, we can extract a sample graph from the ensemble by running a Bernoulli trial for each pair of vertices to connect $i$ and $j$ with probability $p_{ij}$ ($a_{ij}=1$) and not connect with probability $1-p_{ij}$ ($a_{ij}=0$). Repeating this last step, we can generate any desired number of networks that, on average, have the same degree sequence as the observed one. Fig.~\ref{fig:degree_ubcm} shows a good agreement between the average degree vs the empirical ones.

Similarly, for the weighted network we have considered the ``undirected enhanced configuration model'' (UECM), where the constraints are the degree and strength sequences. Again, the constraints are met on average over the network samples (i.e., {\it canonical ensemble}). In this case, the probability $p_{ij}$ is given by
\begin{equation}
p_{ij} \equiv \frac{x_i x_j y_i y_j}{1-y_i y_j+x_i x_j y_i y_j}
\end{equation}

\noindent
and the ${\bf x}$ and ${\bf y}$ vectors can be computed, again, by either maximizing the log-likelihood
\begin{multline}
\lambda({\bf x},{\bf y}) \equiv \sum_i \left[ k_i({\bf W}) \ln x_i + s_i({\bf W}) \ln y_i\right]
+ \sum_i \sum_{j<i} \ln \frac{1-y_i y_j}{1-y_iy_j+x_ix_jy_iy_j}
\end{multline}
\noindent
where ${\bf W}$ represents in this case the adjacency matrix of the weighted graph, or by solving the $2N$ equations

\begin{equation}
\langle k_i\rangle= \sum_{j \neq i} p_{ij} = k_i \left({\bf W} \right) \quad \forall i, 
\end{equation}
\begin{equation}
\langle s_i\rangle= \sum_{j \neq i} \frac{p_{ij}}{1-y_i y_j} = s_i \left({\bf W} \right) \quad\forall i, 
\end{equation}
\noindent where $k_i({\bf W})$ and $s_i({\bf W})$ are, respectively, the observed degree and strength of node $i$ and $\langle k_i\rangle$ and $\langle s_i\rangle$ are the ensemble averages. 

Thus, solving the above equations, the probabilities of generating a link of weight $w$ between any pair of nodes $i$ and $j$ is given by
\begin{equation}
q_{ij}= \begin{cases}1 - p_{ij} , & \mbox{if } w=0,\\ p_{ij} (y_i y_j)^{w-1}(1-y_i y_j), & \mbox{if } w>0.  \end{cases}
\end{equation}

\begin{figure}[!h]
\begin{center}
\includegraphics[width=\columnwidth]{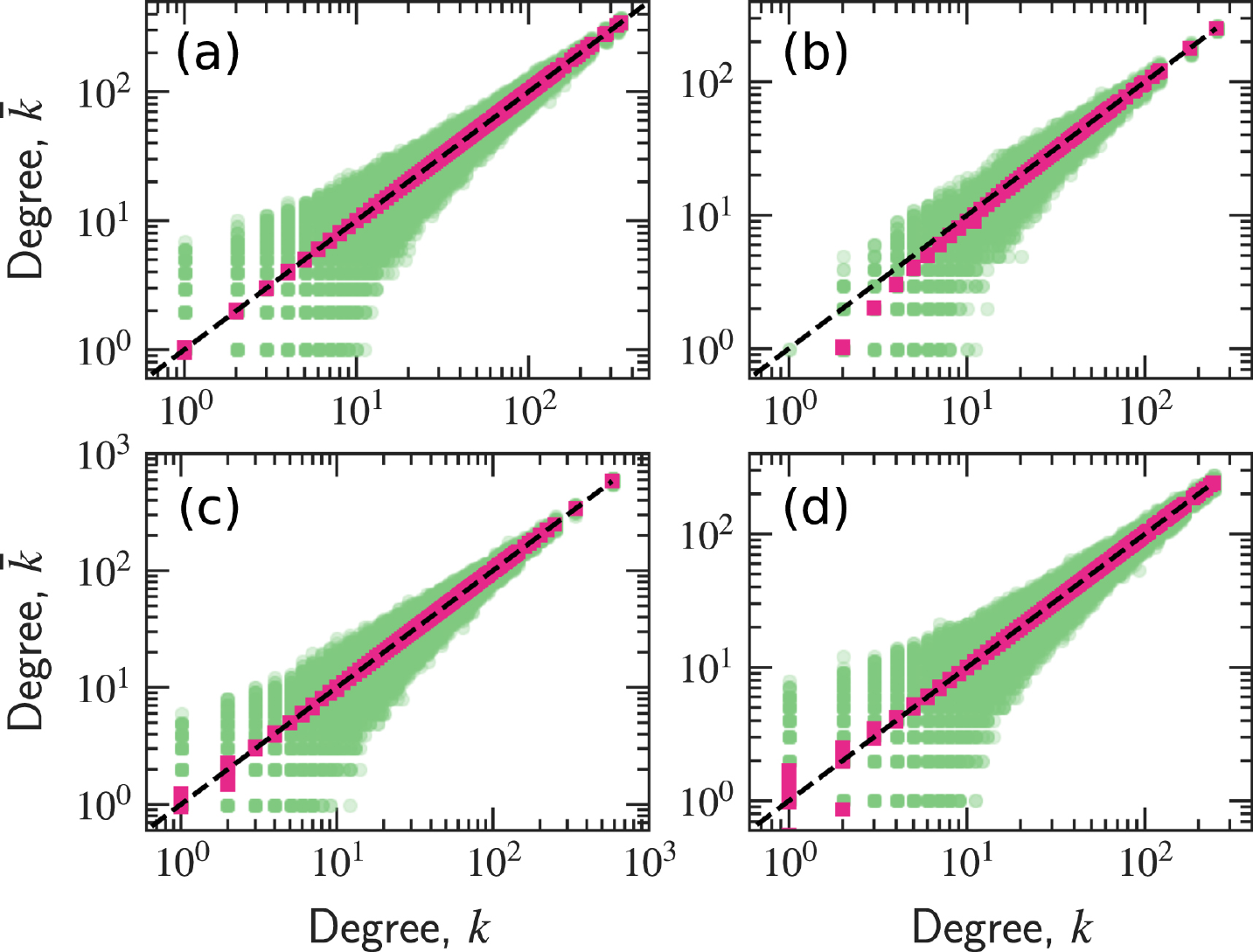} 
\end{center}
\caption{Samples of networks using undirected enhanced configuration model (UECM) given their degree and strength sequences. In each plot, the green dots show the measured degree $\bar{k_i}$ in each sample versus the degree $k_i$ in the observed network. The pink squared-dots represents the average degree $\langle k_i \rangle$ over the $10,000$ networks of the ensemble versus the empirical degree $k_i$. The dashed line is a straight line with slope 1. (a) Brazilian buses transportation network, (b) British buses transportation network, (c) Spanish buses transportation network, and (d) worldwide air transportation network. }
\label{fig:degree_uecm}
\end{figure}

\begin{figure}[!h]
\begin{center}
\includegraphics[width=\columnwidth]{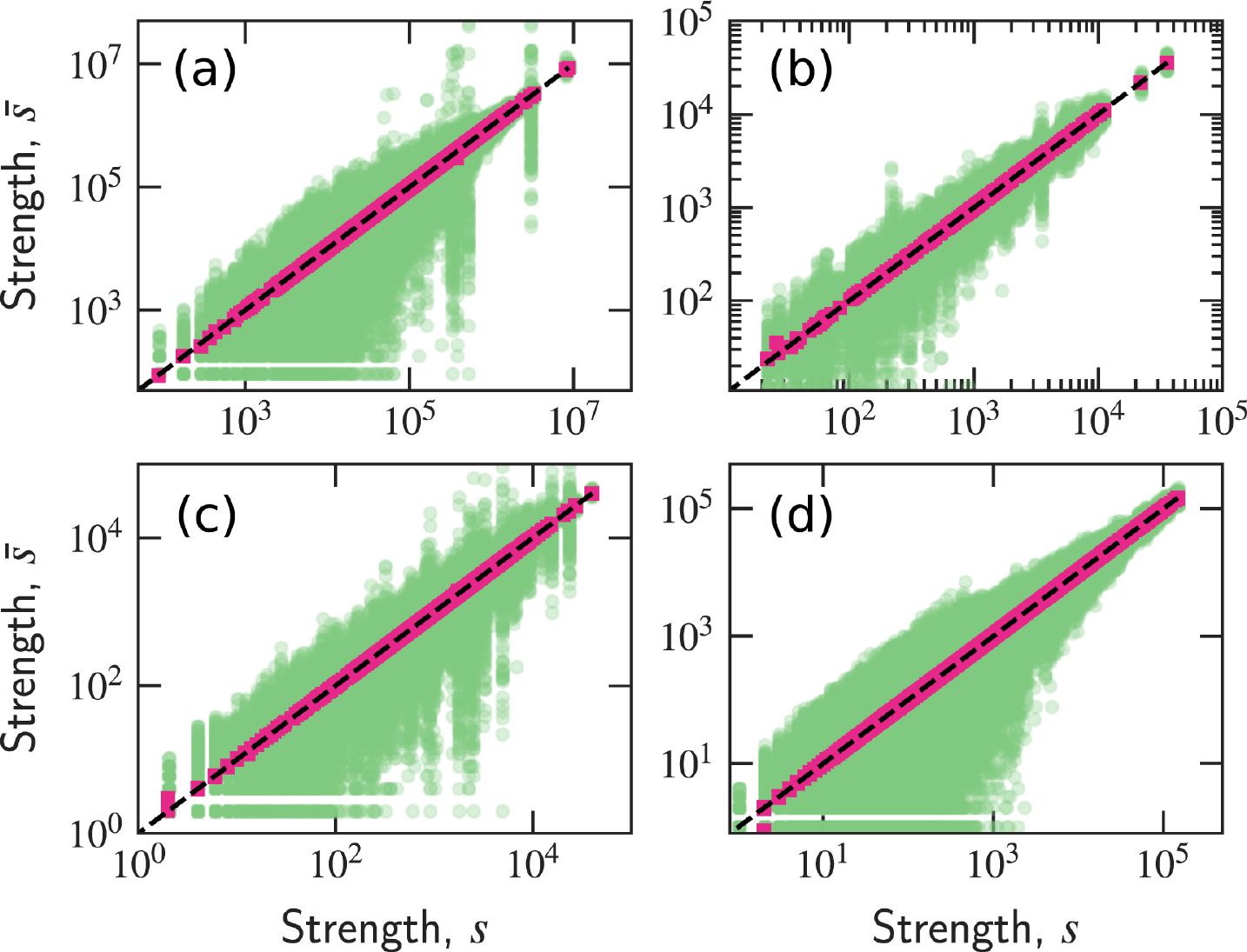} 
\end{center}
\caption{Samples of networks using undirected enhanced configuration model (UECM) given their degree and strength sequences. In each plot, the green dots show the measured strength $\bar{s_i}$ in each sample versus the strength $s_i$ in the observed network. The pink squared-dots represents the average strength $\langle s_i \rangle$ over the $10,000$ networks of the ensemble versus the empirical strength $s_i$. The dashed line is a straight line with slope 1. (a) Brazilian buses transportation network, (b) British buses transportation network, (c) Spanish buses transportation network, and (d) worldwide air transportation network.  }
\label{fig:strength_uecm}
\end{figure}

Figs.~\ref{fig:degree_uecm} and~\ref{fig:strength_uecm} show, respectively,  the average degree and strength over the ensemble generated by the UBCM method compared to the empirical observations. 

{\bf Detecting anomalies.} To detect the anomaly in betweenness centrality versus degree, we have calculated these quantities for each node over a $10,000$ ensemble of synthetic networks considering the appropriate null models. For every node, we approximated the distribution of $k$ and $b$ by a multivariate Gaussian distribution and computed the fraction of nodes that lie outside the $95\%$ confidence interval for the null model.

{\bf Multivariate Gaussian fitting.} For each node, we approximated the joint distribution of betweenness centrality and degree (or strength) by a multivariate Gaussian, that is, 
\begin{equation}
\mathcal{N}(\mathbf x,\{\boldsymbol\mu,\boldsymbol\Sigma\}) = \frac{\exp\left(-\frac{1}{2}({\mathbf   x}-{\boldsymbol\mu})^\mathrm{T}{\boldsymbol\Sigma}^{-1}({\mathbf x}-{\boldsymbol\mu})\right)}{\sqrt{(2\pi)^2|\boldsymbol\Sigma|}}
\end{equation}
where $\mathbf x = (k,b)^T$, 
\begin{equation}
\boldsymbol\mu = \begin{pmatrix} 
                 \mu_k \\ \mu_b 
                 \end{pmatrix}, 
\end{equation}
is the mean, and
\begin{equation}
\boldsymbol\Sigma = \rho \begin{pmatrix} 
                         \sigma_{kk} & \sigma_{kb}\\
                         \sigma_{kb}  & \sigma_{bb} 
                        \end{pmatrix},
\end{equation}
is the covariance matrix, where $\rho$ is the correlation between $k$ and $b$. Thus, the line enclosing $95\%$ of the probability mass for the null model is a ellipsoid (under a rotated coordinate system) with radii given by the eigenvalues $\sqrt{\lambda_1}$ and $\sqrt{\lambda_2}$ of the scaled covariance matrix $s \dot \Sigma$, where $s=-2\log (1-p)$ and $p$ is the confidence probability that the null hypothesis is true.

\section*{Acknowledgments}
LGAA and AA contributed equally to this work. LGAA acknowledges FAPESP (2016/16987-7) for partial financial support. AA acknowledges the support of the FPI doctoral fellowship from MINECO and its mobility scheme. FAR acknowledges the Leverhulme Trust, CNPq (305940/2010-4) and FAPESP (2016/25682-5 and 2013/07375-0) for the financial support given to this research. YM acknowledges partial support from the Government of Arag\'on, Spain through grant E36-17R (FENOL), and by MINECO and FEDER funds (FIS2017-87519-P). LANA thanks the John and Leslie McQuown Gift.

\end{document}